\documentclass[
superscriptaddress,
 amsmath,amssymb,
 aps,
reprint
]{revtex4-1}

\usepackage{color}
\usepackage{graphicx}
\usepackage{dcolumn}
\usepackage{bm}
\usepackage{epstopdf}
\usepackage{gensymb}

\usepackage{float}

\begin{document}


\title{Selective area epitaxy of GaAs films using patterned graphene on Ge}
\author{Zheng Hui Lim}

\author{Sebastian Manzo}

\author{Patrick J. Strohbeen}

\author{Vivek Saraswat}

\author{Michael S. Arnold}

\author{Jason K. Kawasaki}
\affiliation{Materials Science and Engineering, University of Wisconsin-Madison, Madison, WI 53706}

\date{\today}

\begin{abstract}

We demonstrate selective area epitaxy of GaAs films using patterned graphene masks on a Ge (001) substrate. The GaAs selectively grows on exposed regions of the Ge substrate, for graphene spacings as large as 10 microns. The selectivity is highly dependent on the growth temperature and annealing time, which we explain in terms of temperature dependent sticking coefficients and surface diffusion. The high nucleation selectivity over several microns sets constraints on experimental realizations of remote epitaxy.

\end{abstract}

\maketitle

\section{Introduction}

Selective area epitaxy is a promising strategy for synthesis of confined quantum devices \cite{gao2014selective, aseev2018selectivity, lee2019selective} and for reducing dislocation densities in highly lattice mismatched systems \cite{ujiie1989epitaxial, fitzgerald1991epitaxial, park2007defect, mcmahon2018perspective}. This growth mode is typically implemented by patterning openings in an inert dielectric mask such as SiO$_2$, on a crystalline substrate. The openings serve as selective nucleation sites for epitaxial film growth, while the mask defines the lateral dimensions and blocks dislocations from the substrate or film/substrate interface. Continued epitaxial lateral overgrowth can be used to produce coalesced planar films \cite{nishinaga1988epitaxial, ujiie1989epitaxial, mcmahon2018perspective, ironside2019high}.

Due to its chemical inertness, fast surface diffusion, and atomic thinness, patterned graphene is an attractive alternative mask material. Early studies demonstrated epitaxial lateral overgrowth of GaAs films on patterned multilayer graphite on a Si substrate, by liquid phase epitaxy (LPE) \cite{zytkiewicz2001control, zytkiewicz1998microscopic}. More recently, monolayer graphene masks have been used to selectively nucleate GaN films at 75 nm openings on a SiC substrate \cite{puybaret2016nanoselective}, for GaN grown by metalorganic chemical vapor deposition (MOCVD). Open questions remain, however, about whether nucleation selectivity can be achieved using graphene over larger length scales, whether selectivity requires chemical precursors or can be achieved using physical vapor deposition techniques like molecular-beam epitaxy (MBE), and how growth kinetics affect the selectivity.

Here we demonstrate that monolayer graphene masks provide a highly selective barrier for selective area epitaxy of MBE-grown GaAs films, using patterned graphene stripes on a Ge (001) substrate. We demonstrate $>99\%$ nucleation selectivity on the exposed Ge regions, for graphene stripe widths and spacings as large as 10 microns. The nucleation selectivity depends strongly on Ga surface diffusion and desorption, which we control experimentally by increasing the growth temperature and periodic anneal time. The high selectivity over a 10 micron scale also places constraints on experimental realizations of ``remote epitaxy'' \cite{kim2017remote}, where the goal is to grow epitaxial films via remote interactions through continuous graphene with no intentional openings.                                                       

\section{Results and Discussion}

\begin{figure}
    \centering
    \includegraphics[width=0.4\textwidth]{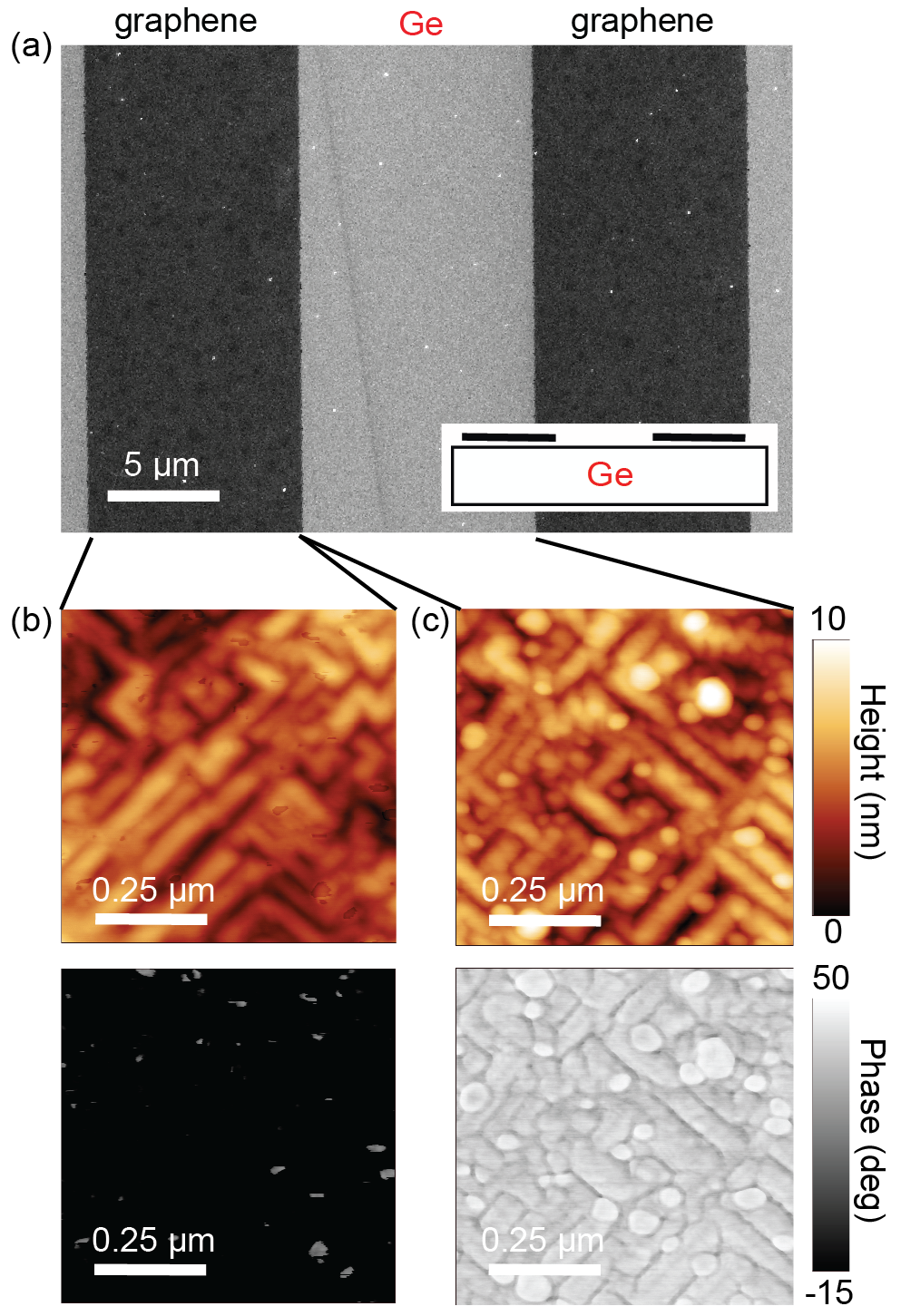}
    \caption{\textbf{Patterned graphene on Ge (001).} (a) Scanning electron micrograph (SEM) image of the graphene mask with 10 $\mu$m feature size under in-lens secondary electron imaging mode. Inset shows the cartoon of graphene mask.
    (b) AFM height (top) and phase (bottom) images of a graphene region. (c) AFM images of an exposed Ge region.}
    \label{pattern}
\end{figure}

To make the patterned graphene on Ge, we first grow continuous monolayer graphene on a Ge (001) substrate by chemical vapor deposition using a methane precursor, following Refs. \cite{kiraly2015, wang2013grge}. We define the stripes by UV photolithography (Shipley 1813 photoresist), etch the exposed graphene regions using an oxygen plasma (50 W), and rinse the remaining photoresist using sequential acetone and isopropanol baths. The resulting patterned template consists of alternating 5-10 microns wide stripes of graphene, separated by 5-10 microns of exposed Ge substrate (Fig. \ref{pattern}a). At this stage, scanning electron microscopy (SEM) confirms that the pattern is translated to the graphene with no obvious long-range tears (Fig. \ref{pattern}a). 

Atomic force microscopy (AFM) images reveal a faceted morphology on both the graphene regions (Fig. \ref{pattern}b) and exposed Ge regions (Fig. \ref{pattern}c). The faceting is consistent with previous studies and is induced by graphene growth \cite{mcelhinny2016graphene}. On the Ge regions we also observe spherical particles that we attribute to Ge that has been oxidized during the oxygen plasma etch. On graphene regions we observe $\sim 10$ nm diameter pinholes. These pinholes appear as bright spots in the AFM phase images, which we attribute to a different elastic modulus of graphene versus the Ge substrate. The pinholes are only faintly visible in the AFM height images. Based on these images we estimate a pinhole concentration of $\sim 40$ $\mu m^{-2}$, which is an order of magnitude smaller than the pinhole concentration for transferred graphene on GaAs (001) after native oxide desorption \cite{manzo2021defect}. We attribute these pinholes to the photolithography and etching processes, since no pinholes were detected by AFM on the graphene/Ge samples before photolithography and etching.

\begin{figure}
    \centering
    \includegraphics[width=0.45\textwidth]{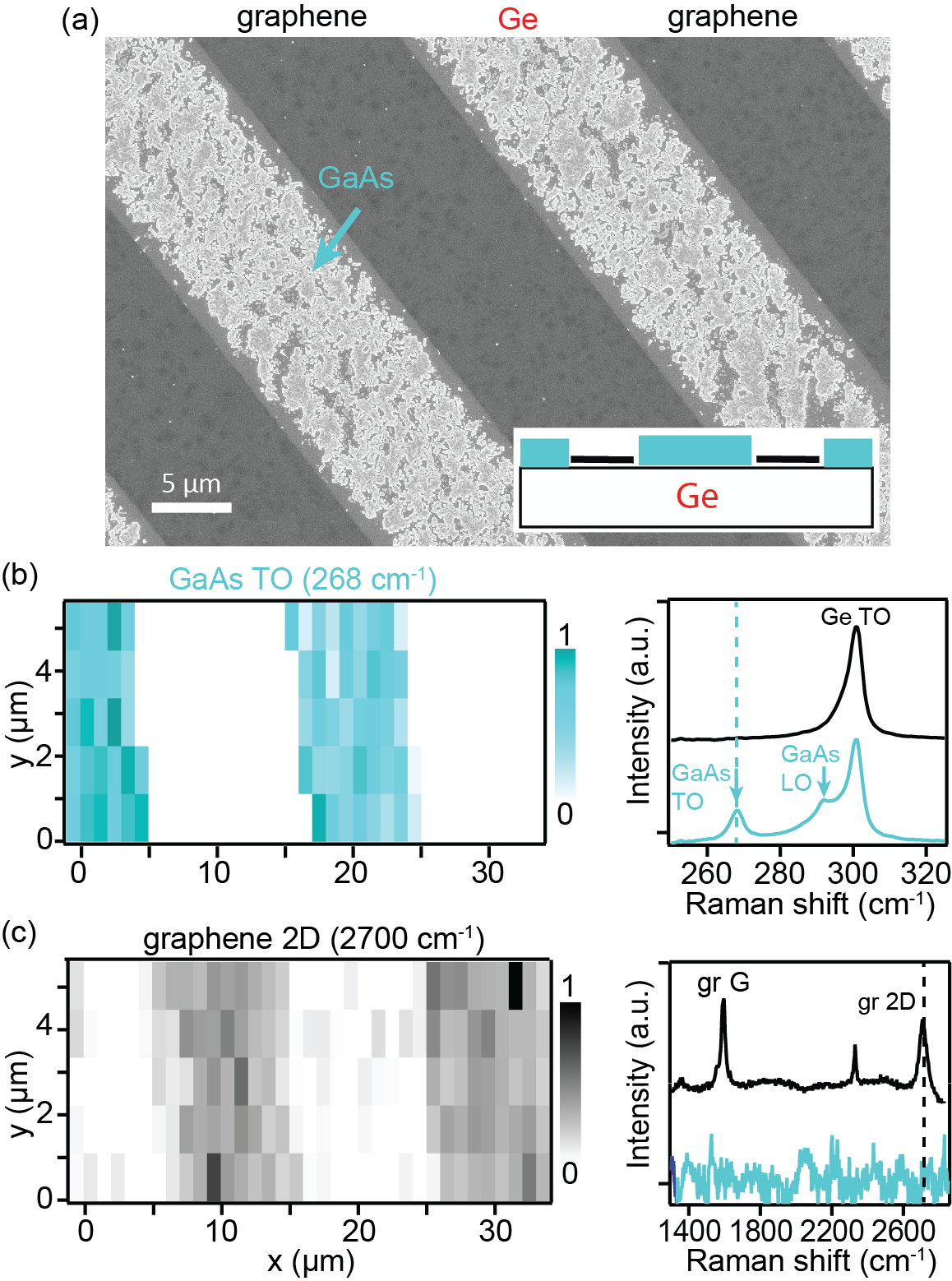}
    \caption{\textbf{GaAs nucleation selectivity at the patterned openings.} (a) Secondary electron SEM image of a GaAs film nucleated on patterned graphene/Ge(001). The stripe pattern consists of 10 microns graphene and 10 microns of exposed Ge. (b) Left: Raman intensity map of the GaAs TO mode. Right: representative point spectra on a GaAs covered region and a graphene region. (c) Raman intensity map of the graphene 2D mode and corresponding spectra.}
    \label{nucleation}
\end{figure}

Prior to GaAs film growth, we etch the exposed Ge oxides in a deionized water bath at 90$^\circ$C. We then dry samples with nitrogen and immediately load them into a high vacuum loadlock ($p< 10^{-8}$ Torr) and outgas them at $150^\circ$C for at least 1 hour to remove organic residuals, followed by an anneal at $650^\circ$C in the ultrahigh vacuum MBE chamber ($p< 3\times10^{-10}$ Torr) to remove the remaining Ge-oxides. GaAs films are grown by MBE using elemental Ga from a standard effusion cell and a mixture of As$_2$/As$_4$ from a thermal cracker cell. Typical Ga fluxes were $2.74 \times 10^{16}$ atoms/(cm$^2$s), calibrated by reflection high energy electron diffraction oscillations.

Fig. \ref{nucleation} shows scanning electron microsocopy (SEM) and Raman spectroscopy for a GaAs film grown at $613^\circ$C by periodic supply of Ga and constant supply of As, on patterned graphene / Ge (001). We observe high nucleation selectivity for GaAs at exposed region of the Ge substrate, and negligible parasitic GaAs nucleation on the graphene mask. In the SEM image (Fig. \ref{nucleation}a), the GaAs appears as coalesced islands with high secondary electron intensity on the Ge regions. We attribute the island morphology to the faceted Ge (001) surface, which is induced by graphene growth (Fig. \ref{pattern}c). Note that CVD graphene on (111) and (110)-oriented Ge is known to produce smoother surfaces \cite{kiraly2015}, free from the facetting of the (001) surface. We expect that GaAs grown on these orientations may produce a smoother film surface morphology.

Raman spectroscopy maps (Horiba Labram, $\lambda=532$ nm) of the GaAs TO and graphene 2D modes confirm that these islands are GaAs film, which grow preferentially on the exposed Ge (Fig. \ref{nucleation}b) and not on the graphene mask (Fig. \ref{nucleation}c). Representative point spectra on the GaAs-covered Ge regions (blue curves) and on the graphene regions (black curves) are shown on the right hand side of Figs \ref{nucleation}(b) and (c). Note that in this scattering geometry, only the GaAs LO mode is symmetry allowed for for a perfect GaAs crystal -- the GaAs TO mode is not symmetry allowed \cite{abstreiter1978raman}. We attribute the presence of a GaAs TO mode to strain or point defects.

\begin{figure*}
    \centering
    \includegraphics[width=0.95\textwidth]{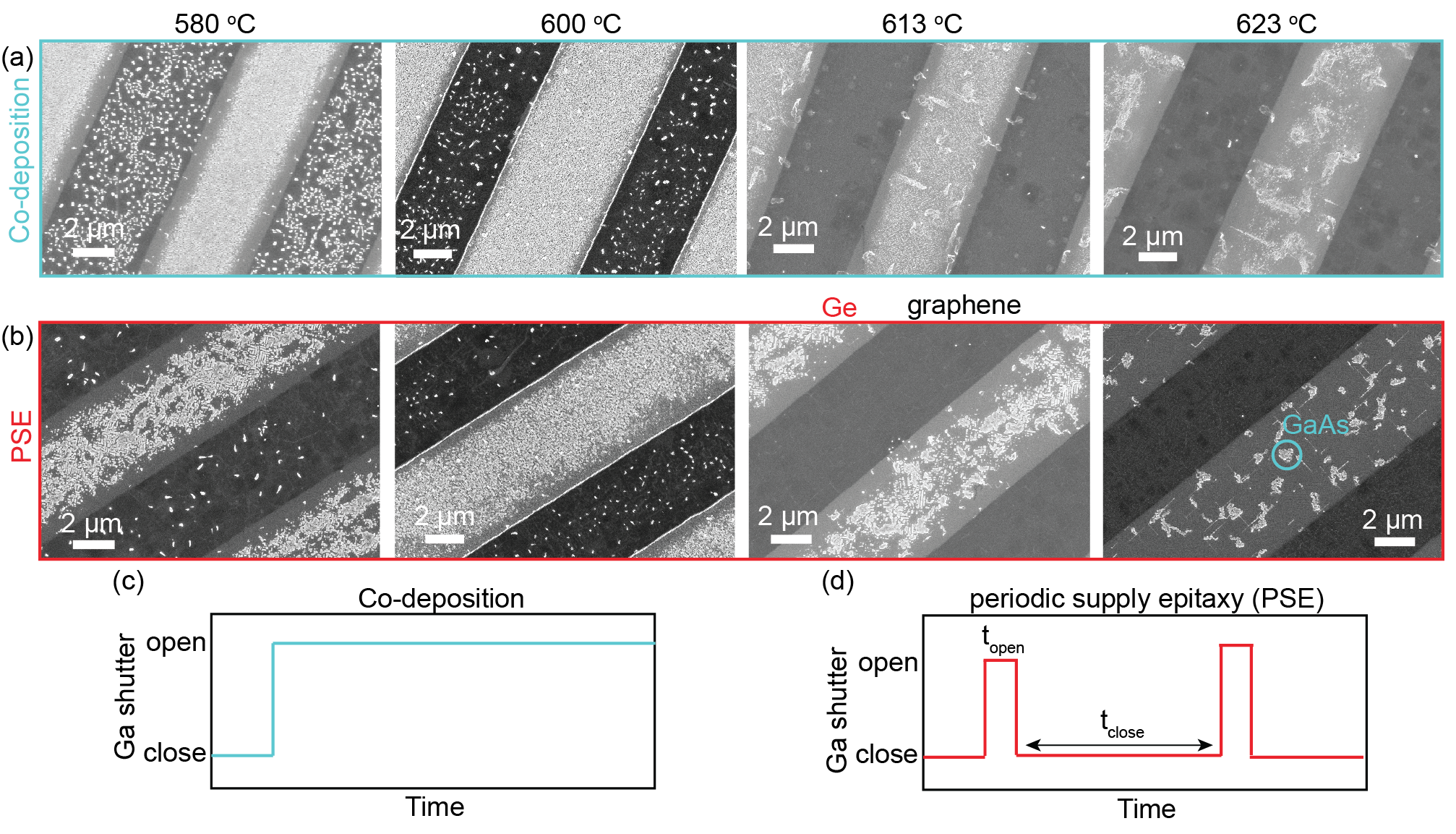}
    \caption{\textbf{Enhancing selectivity with growth temperature and periodic supply epitaxy (PSE).} (a) SEM images of GaAs grown by co-deposition, for increasing growth temperature. (b) SEM images for samples grown by PSE.
    (c) Schematic of co-deposition. The Ga shutter is kept open for the entire growth. (d) Schematic of PSE. The Ga shutter is periodically shuttered open and closed.}
    \label{temperature}
\end{figure*}

\begin{figure}
    \centering
    \includegraphics[width=0.45\textwidth]{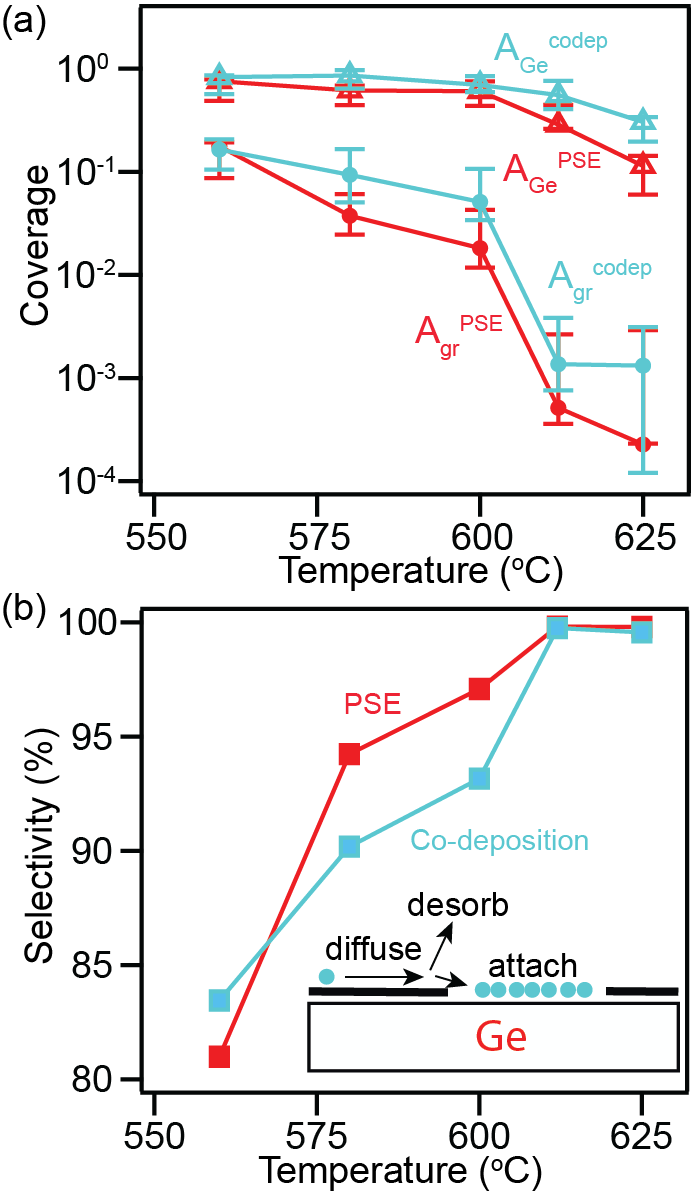}
    \caption{\textbf{Quantification of selectivity.} (a) Percent area coverage of GaAs film nucleated on exposed Ge regions ($A_{Ge}$) and on graphene regions ($A_{gr}$), for growth by co-deposition and by periodic supply epitaxy. The percent area coverage was determined from the SEM images.
    (b) Nucleation selectivity $S = A_{Ge} / (A_{Ge} + A_{gr})$ on the exposed Ge regions.}
    \label{selectivity_quant}
\end{figure}

We find that the selectivity for GaAs growth on Ge rather than the graphene mask is highly dependent on the growth temperature and annealing. Figure \ref{temperature}(a) shows SEM images of GaAs films grown by codeposition of Ga and As, at varying substrate temperature. All samples were exposed to a total flux of $4.56 \times 10^{14}$ Ga atoms/cm$^2$, which would correspond to a GaAs film thickness of 14 nm for a Ga sticking coefficient of 1. In these images the GaAs corresponds to the light islands, the graphene corresponds to the dark stripes, and the Ge corresponds to lighter stripes. At a growth temperature of $580^\circ$C, the concentration of parasitic nucleated islands on graphene is $\sim 10$ $\mu$m$^{-2}$, which is slightly less than the $\sim 40$ $\mu$m$^{-2}$ concentration of graphene pinholes before GaAs growth (Fig. \ref{pattern}b). We hypothesize that the parasitic GaAs growth on graphene is caused by nucleation in the pinholes. Similar nucleation at pinholes is observed for GaAs films grown on unpatterned graphene/GaAs (001) \cite{manzo2021defect}. Further optimization is required to reduce the graphene pinhole density created during lithography and etching.

With increasing growth temperature, the amount of GaAs nucleated on both graphene and Ge regions decreases, i.e. the sticking coefficient decreases. The rate of GaAs sticking on graphene decreases more rapidly than on Ge, such that for a growth temperature of $613^\circ$C there is negligible GaAs growth on graphene and highly preferred growth on Ge. We quantify these effects in Fig. \ref{selectivity_quant}. Here, $A_{Ge}^{codep}$ is the area fraction of GaAs on the exposed Ge regions and $A_{gr}^{codep}$ is the area fraction of GaAs on graphene (Supplemental Information). We define the selectivity as $S = A_{Ge}/(A_{gr}+A_{Ge})$. At growth temperatures greater than $610^\circ$C we obtain a selectivity $S >99\%$.

We further enhance the nucleation selectivity by employing periodic supply epitaxy (PSE). Here, we fix the As shutter open and periodically modulate the Ga shutter between the open position (GaAs growth) and the closed position (annealing, Fig. \ref{temperature}d). Fig. \ref{temperature}b shows SEM images of GaAs films grown by PSE, where we let $t_{growth} = $ 1 minute and $t_{close}=$ 6 minutes, for a total of 15 cycles. We find that PSE decreases the total sticking coefficient (Fig. \ref{selectivity_quant}a) and increases the selectivity for GaAs nucleation on Ge compared to on graphene (Fig. \ref{selectivity_quant}b).

We understand effects of temperature and anneal time in terms of the microscopic processes sketched in Fig. \ref{selectivity_quant}b insert. High selectivity results from a combination of fast surface diffusion on graphene and high desorption rate from graphene, compared to Ge which has a sticking coefficient closer to 1. Upon arrival at the surface, Ga adatoms diffuse over a characteristic length $\lambda$. Ga adatoms will then either desorb into the vapor or attach to the surface to form GaAs. With increasing temperature, the desorption rate of Ga from graphene decreases dramatically, as observed by the factor of $10^3$ drop in GaAs surface coverage on graphene ($A_{gr}$) from 560 to 625$^\circ$C (Fig. \ref{selectivity_quant}a). In contrast, for the same range of temperatures the GaAs coverage on Ge ($A_{Ge}$) decreases by less than a factor of 10. Surface diffusion also increases with temperature, resulting in an increased probability for Ga to diffuse and attach to reactive areas of the exposed Ge substrate. These combined factors explain the general observation that increasing the temperature increases the selective growth on Ge. Increasing the anneal time by periodic supply epitaxy plays a similar role as temperature, by increasing the time for desorption and increasing the diffusion length $\lambda$. These general trends are consistent with selective area growth using conventional SiO$_2$ masks \cite{nishinaga1988epitaxial}.

The high selectivity and long diffusion length also set constraints for experimental realizations of ``remote epitaxy'' on unpatterned graphene. ``Remote epitaxy'' refers to the epitaxial growth of a film on a continuous graphene-terminated substrate, with no intentional openings. In this growth mode, epitaxial registry between film and substrate is thought to occur via ``remote'' interactions that permeate through graphene. However, pinholes or openings in the graphene can serve as alternate nucleation sites, followed by lateral overgrowth \cite{manzo2021defect} or growth by intercalation \cite{briggs2020atomically}. In the present work, our results show that GaAs can nucleate at graphene openings over length scales as large as 10 microns. Experimental realizations of remote epitaxy may require graphene defect control over a length scale of several microns, in order to clearly distinguish a ``remote'' mechanism from a selective area and lateral epitaxy mechanism.

\section{Conclusions}

We demonstrated selective area epitaxy of GaAs using a patterned graphene mask on Ge (001), over a length scale of 10 microns. The selectivity is comparable to previous demonstrations of selective area epitaxy by MBE using SiO$_2$ masks, where smaller spacings of a 100 nm to 2 microns are more typical \cite{ironside2019high, lee2002selective, bacchin2000fabrication}. The main limiting feature in the current work is the rough surface morphology of GaAs on Ge (001), which we attribute to graphene-induced faceting of the Ge (001) surface. We expect that growth on graphene/Ge in (111) or (110) orientation may be a path for smoother films, since these orientations do not produce the large graphene-induced faceting. Further improvements in lithography and etching may also be required to reduce the graphene pinholes and enhance the selectivity for growth at temperatures below $600^\circ$C.

\section{Acknowledgments}

This work was primarily supported by the Defense Advanced Research Projects Administration (DARPA Grant number D19AP00088). Graphene synthesis and characterization are supported by the U.S. Department of Energy, Office of Science, Basic Energy Sciences, under award no. DE-SC0016007. GaAs film characterization was supported by the NSF Division of Materials Research through the University of Wisconsin Materials Research Science and Engineering Center (Grant No. DMR-1720415) and the CAREER program (DMR-1752797). We gratefully acknowledge the use of Raman and electron microscopy facilities supported by the NSF through the University of Wisconsin Materials Research Science and Engineering Center under Grant No. DMR-1720415.

\bibliographystyle{apsrev}
\bibliography{reference}

\begin{thebibliography}{21}
\expandafter\ifx\csname natexlab\endcsname\relax\def\natexlab#1{#1}\fi
\expandafter\ifx\csname bibnamefont\endcsname\relax
  \def\bibnamefont#1{#1}\fi
\expandafter\ifx\csname bibfnamefont\endcsname\relax
  \def\bibfnamefont#1{#1}\fi
\expandafter\ifx\csname citenamefont\endcsname\relax
  \def\citenamefont#1{#1}\fi
\expandafter\ifx\csname url\endcsname\relax
  \def\url#1{\texttt{#1}}\fi
\expandafter\ifx\csname urlprefix\endcsname\relax\def\urlprefix{URL }\fi
\providecommand{\bibinfo}[2]{#2}
\providecommand{\eprint}[2][]{\url{#2}}

\bibitem[{\citenamefont{Gao et~al.}(2014)\citenamefont{Gao, Saxena, Wang, Fu,
  Mokkapati, Guo, Li, Wong-Leung, Caroff, Tan et~al.}}]{gao2014selective}
\bibinfo{author}{\bibfnamefont{Q.}~\bibnamefont{Gao}},
  \bibinfo{author}{\bibfnamefont{D.}~\bibnamefont{Saxena}},
  \bibinfo{author}{\bibfnamefont{F.}~\bibnamefont{Wang}},
  \bibinfo{author}{\bibfnamefont{L.}~\bibnamefont{Fu}},
  \bibinfo{author}{\bibfnamefont{S.}~\bibnamefont{Mokkapati}},
  \bibinfo{author}{\bibfnamefont{Y.}~\bibnamefont{Guo}},
  \bibinfo{author}{\bibfnamefont{L.}~\bibnamefont{Li}},
  \bibinfo{author}{\bibfnamefont{J.}~\bibnamefont{Wong-Leung}},
  \bibinfo{author}{\bibfnamefont{P.}~\bibnamefont{Caroff}},
  \bibinfo{author}{\bibfnamefont{H.~H.} \bibnamefont{Tan}},
  \bibnamefont{et~al.}, \bibinfo{journal}{Nano letters}
  \textbf{\bibinfo{volume}{14}}, \bibinfo{pages}{5206} (\bibinfo{year}{2014}).

\bibitem[{\citenamefont{Aseev et~al.}(2018)\citenamefont{Aseev, Fursina,
  Boekhout, Krizek, Sestoft, Borsoi, Heedt, Wang, Binci, Mart{\'\i}-S{\'a}nchez
  et~al.}}]{aseev2018selectivity}
\bibinfo{author}{\bibfnamefont{P.}~\bibnamefont{Aseev}},
  \bibinfo{author}{\bibfnamefont{A.}~\bibnamefont{Fursina}},
  \bibinfo{author}{\bibfnamefont{F.}~\bibnamefont{Boekhout}},
  \bibinfo{author}{\bibfnamefont{F.}~\bibnamefont{Krizek}},
  \bibinfo{author}{\bibfnamefont{J.~E.} \bibnamefont{Sestoft}},
  \bibinfo{author}{\bibfnamefont{F.}~\bibnamefont{Borsoi}},
  \bibinfo{author}{\bibfnamefont{S.}~\bibnamefont{Heedt}},
  \bibinfo{author}{\bibfnamefont{G.}~\bibnamefont{Wang}},
  \bibinfo{author}{\bibfnamefont{L.}~\bibnamefont{Binci}},
  \bibinfo{author}{\bibfnamefont{S.}~\bibnamefont{Mart{\'\i}-S{\'a}nchez}},
  \bibnamefont{et~al.}, \bibinfo{journal}{Nano letters}
  \textbf{\bibinfo{volume}{19}}, \bibinfo{pages}{218} (\bibinfo{year}{2018}).

\bibitem[{\citenamefont{Lee et~al.}(2019)\citenamefont{Lee, Choi, Pendharkar,
  Pennachio, Markman, Seas, Koelling, Verheijen, Casparis, Petersson
  et~al.}}]{lee2019selective}
\bibinfo{author}{\bibfnamefont{J.~S.} \bibnamefont{Lee}},
  \bibinfo{author}{\bibfnamefont{S.}~\bibnamefont{Choi}},
  \bibinfo{author}{\bibfnamefont{M.}~\bibnamefont{Pendharkar}},
  \bibinfo{author}{\bibfnamefont{D.~J.} \bibnamefont{Pennachio}},
  \bibinfo{author}{\bibfnamefont{B.}~\bibnamefont{Markman}},
  \bibinfo{author}{\bibfnamefont{M.}~\bibnamefont{Seas}},
  \bibinfo{author}{\bibfnamefont{S.}~\bibnamefont{Koelling}},
  \bibinfo{author}{\bibfnamefont{M.~A.} \bibnamefont{Verheijen}},
  \bibinfo{author}{\bibfnamefont{L.}~\bibnamefont{Casparis}},
  \bibinfo{author}{\bibfnamefont{K.~D.} \bibnamefont{Petersson}},
  \bibnamefont{et~al.}, \bibinfo{journal}{Physical Review Materials}
  \textbf{\bibinfo{volume}{3}}, \bibinfo{pages}{084606} (\bibinfo{year}{2019}).

\bibitem[{\citenamefont{Ujiie and Nishinaga}(1989)}]{ujiie1989epitaxial}
\bibinfo{author}{\bibfnamefont{Y.}~\bibnamefont{Ujiie}} \bibnamefont{and}
  \bibinfo{author}{\bibfnamefont{T.}~\bibnamefont{Nishinaga}},
  \bibinfo{journal}{Japanese Journal of Applied Physics}
  \textbf{\bibinfo{volume}{28}}, \bibinfo{pages}{L337} (\bibinfo{year}{1989}).

\bibitem[{\citenamefont{Fitzgerald and Chand}(1991)}]{fitzgerald1991epitaxial}
\bibinfo{author}{\bibfnamefont{E.}~\bibnamefont{Fitzgerald}} \bibnamefont{and}
  \bibinfo{author}{\bibfnamefont{N.}~\bibnamefont{Chand}},
  \bibinfo{journal}{Journal of electronic materials}
  \textbf{\bibinfo{volume}{20}}, \bibinfo{pages}{839} (\bibinfo{year}{1991}).

\bibitem[{\citenamefont{Park et~al.}(2007)\citenamefont{Park, Bai, Curtin,
  Adekore, Carroll, and Lochtefeld}}]{park2007defect}
\bibinfo{author}{\bibfnamefont{J.-S.} \bibnamefont{Park}},
  \bibinfo{author}{\bibfnamefont{J.}~\bibnamefont{Bai}},
  \bibinfo{author}{\bibfnamefont{M.}~\bibnamefont{Curtin}},
  \bibinfo{author}{\bibfnamefont{B.}~\bibnamefont{Adekore}},
  \bibinfo{author}{\bibfnamefont{M.}~\bibnamefont{Carroll}}, \bibnamefont{and}
  \bibinfo{author}{\bibfnamefont{A.}~\bibnamefont{Lochtefeld}},
  \bibinfo{journal}{Applied Physics Letters} \textbf{\bibinfo{volume}{90}},
  \bibinfo{pages}{052113} (\bibinfo{year}{2007}).

\bibitem[{\citenamefont{McMahon et~al.}(2018)\citenamefont{McMahon, Vaisman,
  Zimmerman, Tamboli, and Warren}}]{mcmahon2018perspective}
\bibinfo{author}{\bibfnamefont{W.~E.} \bibnamefont{McMahon}},
  \bibinfo{author}{\bibfnamefont{M.}~\bibnamefont{Vaisman}},
  \bibinfo{author}{\bibfnamefont{J.~D.} \bibnamefont{Zimmerman}},
  \bibinfo{author}{\bibfnamefont{A.~C.} \bibnamefont{Tamboli}},
  \bibnamefont{and} \bibinfo{author}{\bibfnamefont{E.~L.}
  \bibnamefont{Warren}}, \bibinfo{journal}{APL Materials}
  \textbf{\bibinfo{volume}{6}}, \bibinfo{pages}{120903} (\bibinfo{year}{2018}).

\bibitem[{\citenamefont{Nishinaga et~al.}(1988)\citenamefont{Nishinaga, Nakano,
  and Zhang}}]{nishinaga1988epitaxial}
\bibinfo{author}{\bibfnamefont{T.}~\bibnamefont{Nishinaga}},
  \bibinfo{author}{\bibfnamefont{T.}~\bibnamefont{Nakano}}, \bibnamefont{and}
  \bibinfo{author}{\bibfnamefont{S.}~\bibnamefont{Zhang}},
  \bibinfo{journal}{Japanese journal of applied physics}
  \textbf{\bibinfo{volume}{27}}, \bibinfo{pages}{L964} (\bibinfo{year}{1988}).

\bibitem[{\citenamefont{Ironside et~al.}(2019)\citenamefont{Ironside, Skipper,
  Leonard, Radulaski, Sarmiento, Dhingra, Lee, Vuckovic, and
  Bank}}]{ironside2019high}
\bibinfo{author}{\bibfnamefont{D.~J.} \bibnamefont{Ironside}},
  \bibinfo{author}{\bibfnamefont{A.~M.} \bibnamefont{Skipper}},
  \bibinfo{author}{\bibfnamefont{T.~A.} \bibnamefont{Leonard}},
  \bibinfo{author}{\bibfnamefont{M.}~\bibnamefont{Radulaski}},
  \bibinfo{author}{\bibfnamefont{T.}~\bibnamefont{Sarmiento}},
  \bibinfo{author}{\bibfnamefont{P.}~\bibnamefont{Dhingra}},
  \bibinfo{author}{\bibfnamefont{M.~L.} \bibnamefont{Lee}},
  \bibinfo{author}{\bibfnamefont{J.}~\bibnamefont{Vuckovic}}, \bibnamefont{and}
  \bibinfo{author}{\bibfnamefont{S.~R.} \bibnamefont{Bank}},
  \bibinfo{journal}{Crystal Growth \& Design} \textbf{\bibinfo{volume}{19}},
  \bibinfo{pages}{3085} (\bibinfo{year}{2019}).

\bibitem[{\citenamefont{Zytkiewicz et~al.}(2001)\citenamefont{Zytkiewicz,
  Domaga{\l}a, and Dobosz}}]{zytkiewicz2001control}
\bibinfo{author}{\bibfnamefont{Z.}~\bibnamefont{Zytkiewicz}},
  \bibinfo{author}{\bibfnamefont{J.}~\bibnamefont{Domaga{\l}a}},
  \bibnamefont{and} \bibinfo{author}{\bibfnamefont{D.}~\bibnamefont{Dobosz}},
  \bibinfo{journal}{Journal of Applied Physics} \textbf{\bibinfo{volume}{90}},
  \bibinfo{pages}{6140} (\bibinfo{year}{2001}).

\bibitem[{\citenamefont{Zytkiewicz et~al.}(1998)\citenamefont{Zytkiewicz,
  Domagala, Dobosz, and Bak-Misiuk}}]{zytkiewicz1998microscopic}
\bibinfo{author}{\bibfnamefont{Z.}~\bibnamefont{Zytkiewicz}},
  \bibinfo{author}{\bibfnamefont{J.}~\bibnamefont{Domagala}},
  \bibinfo{author}{\bibfnamefont{D.}~\bibnamefont{Dobosz}}, \bibnamefont{and}
  \bibinfo{author}{\bibfnamefont{J.}~\bibnamefont{Bak-Misiuk}},
  \bibinfo{journal}{Journal of applied physics} \textbf{\bibinfo{volume}{84}},
  \bibinfo{pages}{6937} (\bibinfo{year}{1998}).

\bibitem[{\citenamefont{Puybaret et~al.}(2016)\citenamefont{Puybaret,
  Patriarche, Jordan, Sundaram, El~Gmili, Salvestrini, Voss, De~Heer, Berger,
  and Ougazzaden}}]{puybaret2016nanoselective}
\bibinfo{author}{\bibfnamefont{R.}~\bibnamefont{Puybaret}},
  \bibinfo{author}{\bibfnamefont{G.}~\bibnamefont{Patriarche}},
  \bibinfo{author}{\bibfnamefont{M.~B.} \bibnamefont{Jordan}},
  \bibinfo{author}{\bibfnamefont{S.}~\bibnamefont{Sundaram}},
  \bibinfo{author}{\bibfnamefont{Y.}~\bibnamefont{El~Gmili}},
  \bibinfo{author}{\bibfnamefont{J.-P.} \bibnamefont{Salvestrini}},
  \bibinfo{author}{\bibfnamefont{P.~L.} \bibnamefont{Voss}},
  \bibinfo{author}{\bibfnamefont{W.~A.} \bibnamefont{De~Heer}},
  \bibinfo{author}{\bibfnamefont{C.}~\bibnamefont{Berger}}, \bibnamefont{and}
  \bibinfo{author}{\bibfnamefont{A.}~\bibnamefont{Ougazzaden}},
  \bibinfo{journal}{Applied Physics Letters} \textbf{\bibinfo{volume}{108}},
  \bibinfo{pages}{103105} (\bibinfo{year}{2016}).

\bibitem[{\citenamefont{Kim et~al.}(2017)\citenamefont{Kim, Cruz, Lee, Alawode,
  Choi, Song, Johnson, Heidelberger, Kong, Choi et~al.}}]{kim2017remote}
\bibinfo{author}{\bibfnamefont{Y.}~\bibnamefont{Kim}},
  \bibinfo{author}{\bibfnamefont{S.~S.} \bibnamefont{Cruz}},
  \bibinfo{author}{\bibfnamefont{K.}~\bibnamefont{Lee}},
  \bibinfo{author}{\bibfnamefont{B.~O.} \bibnamefont{Alawode}},
  \bibinfo{author}{\bibfnamefont{C.}~\bibnamefont{Choi}},
  \bibinfo{author}{\bibfnamefont{Y.}~\bibnamefont{Song}},
  \bibinfo{author}{\bibfnamefont{J.~M.} \bibnamefont{Johnson}},
  \bibinfo{author}{\bibfnamefont{C.}~\bibnamefont{Heidelberger}},
  \bibinfo{author}{\bibfnamefont{W.}~\bibnamefont{Kong}},
  \bibinfo{author}{\bibfnamefont{S.}~\bibnamefont{Choi}}, \bibnamefont{et~al.},
  \bibinfo{journal}{Nature} \textbf{\bibinfo{volume}{544}},
  \bibinfo{pages}{340} (\bibinfo{year}{2017}).

\bibitem[{\citenamefont{Kiraly et~al.}(2015)\citenamefont{Kiraly, Jacobberger,
  Mannix, Campbell, Bedzyk, Arnold, Hersam, and Guisinger}}]{kiraly2015}
\bibinfo{author}{\bibfnamefont{B.}~\bibnamefont{Kiraly}},
  \bibinfo{author}{\bibfnamefont{R.~M.} \bibnamefont{Jacobberger}},
  \bibinfo{author}{\bibfnamefont{A.~J.} \bibnamefont{Mannix}},
  \bibinfo{author}{\bibfnamefont{G.~P.} \bibnamefont{Campbell}},
  \bibinfo{author}{\bibfnamefont{M.~J.} \bibnamefont{Bedzyk}},
  \bibinfo{author}{\bibfnamefont{M.~S.} \bibnamefont{Arnold}},
  \bibinfo{author}{\bibfnamefont{M.~C.} \bibnamefont{Hersam}},
  \bibnamefont{and} \bibinfo{author}{\bibfnamefont{N.~P.}
  \bibnamefont{Guisinger}}, \bibinfo{journal}{Nano Lett.}
  \textbf{\bibinfo{volume}{15}}, \bibinfo{pages}{7414} (\bibinfo{year}{2015}).

\bibitem[{\citenamefont{Wang et~al.}(2013)\citenamefont{Wang, Zhang, Zhu, Ding,
  Jiang, Qinglei, Liu, Xie, Chu, Di et~al.}}]{wang2013grge}
\bibinfo{author}{\bibfnamefont{G.}~\bibnamefont{Wang}},
  \bibinfo{author}{\bibfnamefont{M.}~\bibnamefont{Zhang}},
  \bibinfo{author}{\bibfnamefont{Y.}~\bibnamefont{Zhu}},
  \bibinfo{author}{\bibfnamefont{G.}~\bibnamefont{Ding}},
  \bibinfo{author}{\bibfnamefont{D.}~\bibnamefont{Jiang}},
  \bibinfo{author}{\bibfnamefont{G.}~\bibnamefont{Qinglei}},
  \bibinfo{author}{\bibfnamefont{S.}~\bibnamefont{Liu}},
  \bibinfo{author}{\bibfnamefont{X.}~\bibnamefont{Xie}},
  \bibinfo{author}{\bibfnamefont{P.~K.} \bibnamefont{Chu}},
  \bibinfo{author}{\bibfnamefont{Z.}~\bibnamefont{Di}}, \bibnamefont{et~al.},
  \bibinfo{journal}{Sci. Rep.} \textbf{\bibinfo{volume}{3}},
  \bibinfo{pages}{2465} (\bibinfo{year}{2013}).

\bibitem[{\citenamefont{McElhinny et~al.}(2016)\citenamefont{McElhinny,
  Jacobberger, Zaug, Arnold, and Evans}}]{mcelhinny2016graphene}
\bibinfo{author}{\bibfnamefont{K.~M.} \bibnamefont{McElhinny}},
  \bibinfo{author}{\bibfnamefont{R.~M.} \bibnamefont{Jacobberger}},
  \bibinfo{author}{\bibfnamefont{A.~J.} \bibnamefont{Zaug}},
  \bibinfo{author}{\bibfnamefont{M.~S.} \bibnamefont{Arnold}},
  \bibnamefont{and} \bibinfo{author}{\bibfnamefont{P.~G.} \bibnamefont{Evans}},
  \bibinfo{journal}{Surface Science} \textbf{\bibinfo{volume}{647}},
  \bibinfo{pages}{90} (\bibinfo{year}{2016}).

\bibitem[{\citenamefont{Manzo et~al.}(2021)\citenamefont{Manzo, Strohbeen, Lim,
  Saraswat, Arnold, and Kawasaki}}]{manzo2021defect}
\bibinfo{author}{\bibfnamefont{S.}~\bibnamefont{Manzo}},
  \bibinfo{author}{\bibfnamefont{P.~J.} \bibnamefont{Strohbeen}},
  \bibinfo{author}{\bibfnamefont{Z.-H.} \bibnamefont{Lim}},
  \bibinfo{author}{\bibfnamefont{V.}~\bibnamefont{Saraswat}},
  \bibinfo{author}{\bibfnamefont{M.~S.} \bibnamefont{Arnold}},
  \bibnamefont{and} \bibinfo{author}{\bibfnamefont{J.~K.}
  \bibnamefont{Kawasaki}}, \bibinfo{journal}{arXiv preprint arXiv:2106.00721}
  (\bibinfo{year}{2021}).

\bibitem[{\citenamefont{Abstreiter et~al.}(1978)\citenamefont{Abstreiter,
  Bauser, Fischer, and Ploog}}]{abstreiter1978raman}
\bibinfo{author}{\bibfnamefont{G.}~\bibnamefont{Abstreiter}},
  \bibinfo{author}{\bibfnamefont{E.}~\bibnamefont{Bauser}},
  \bibinfo{author}{\bibfnamefont{A.}~\bibnamefont{Fischer}}, \bibnamefont{and}
  \bibinfo{author}{\bibfnamefont{K.}~\bibnamefont{Ploog}},
  \bibinfo{journal}{Applied physics} \textbf{\bibinfo{volume}{16}},
  \bibinfo{pages}{345} (\bibinfo{year}{1978}).

\bibitem[{\citenamefont{Briggs et~al.}(2020)\citenamefont{Briggs, Bersch, Wang,
  Jiang, Koch, Nayir, Wang, Kolmer, Ko, Duran et~al.}}]{briggs2020atomically}
\bibinfo{author}{\bibfnamefont{N.}~\bibnamefont{Briggs}},
  \bibinfo{author}{\bibfnamefont{B.}~\bibnamefont{Bersch}},
  \bibinfo{author}{\bibfnamefont{Y.}~\bibnamefont{Wang}},
  \bibinfo{author}{\bibfnamefont{J.}~\bibnamefont{Jiang}},
  \bibinfo{author}{\bibfnamefont{R.~J.} \bibnamefont{Koch}},
  \bibinfo{author}{\bibfnamefont{N.}~\bibnamefont{Nayir}},
  \bibinfo{author}{\bibfnamefont{K.}~\bibnamefont{Wang}},
  \bibinfo{author}{\bibfnamefont{M.}~\bibnamefont{Kolmer}},
  \bibinfo{author}{\bibfnamefont{W.}~\bibnamefont{Ko}},
  \bibinfo{author}{\bibfnamefont{A.~D. L.~F.} \bibnamefont{Duran}},
  \bibnamefont{et~al.}, \bibinfo{journal}{Nature materials}
  \textbf{\bibinfo{volume}{19}}, \bibinfo{pages}{637} (\bibinfo{year}{2020}).

\bibitem[{\citenamefont{Lee et~al.}(2002)\citenamefont{Lee, Malloy, Dawson, and
  Brueck}}]{lee2002selective}
\bibinfo{author}{\bibfnamefont{S.}~\bibnamefont{Lee}},
  \bibinfo{author}{\bibfnamefont{K.}~\bibnamefont{Malloy}},
  \bibinfo{author}{\bibfnamefont{L.}~\bibnamefont{Dawson}}, \bibnamefont{and}
  \bibinfo{author}{\bibfnamefont{S.}~\bibnamefont{Brueck}},
  \bibinfo{journal}{Journal of applied physics} \textbf{\bibinfo{volume}{92}},
  \bibinfo{pages}{6567} (\bibinfo{year}{2002}).

\bibitem[{\citenamefont{Bacchin and Nishinaga}(2000)}]{bacchin2000fabrication}
\bibinfo{author}{\bibfnamefont{G.}~\bibnamefont{Bacchin}} \bibnamefont{and}
  \bibinfo{author}{\bibfnamefont{T.}~\bibnamefont{Nishinaga}},
  \bibinfo{journal}{Journal of crystal growth} \textbf{\bibinfo{volume}{211}},
  \bibinfo{pages}{389} (\bibinfo{year}{2000}).

\end{thebibliography}

\section{Supporting Information}

\begin{figure*}
    \renewcommand\thefigure{S-1}
    \centering
    \includegraphics[width=0.95\textwidth]{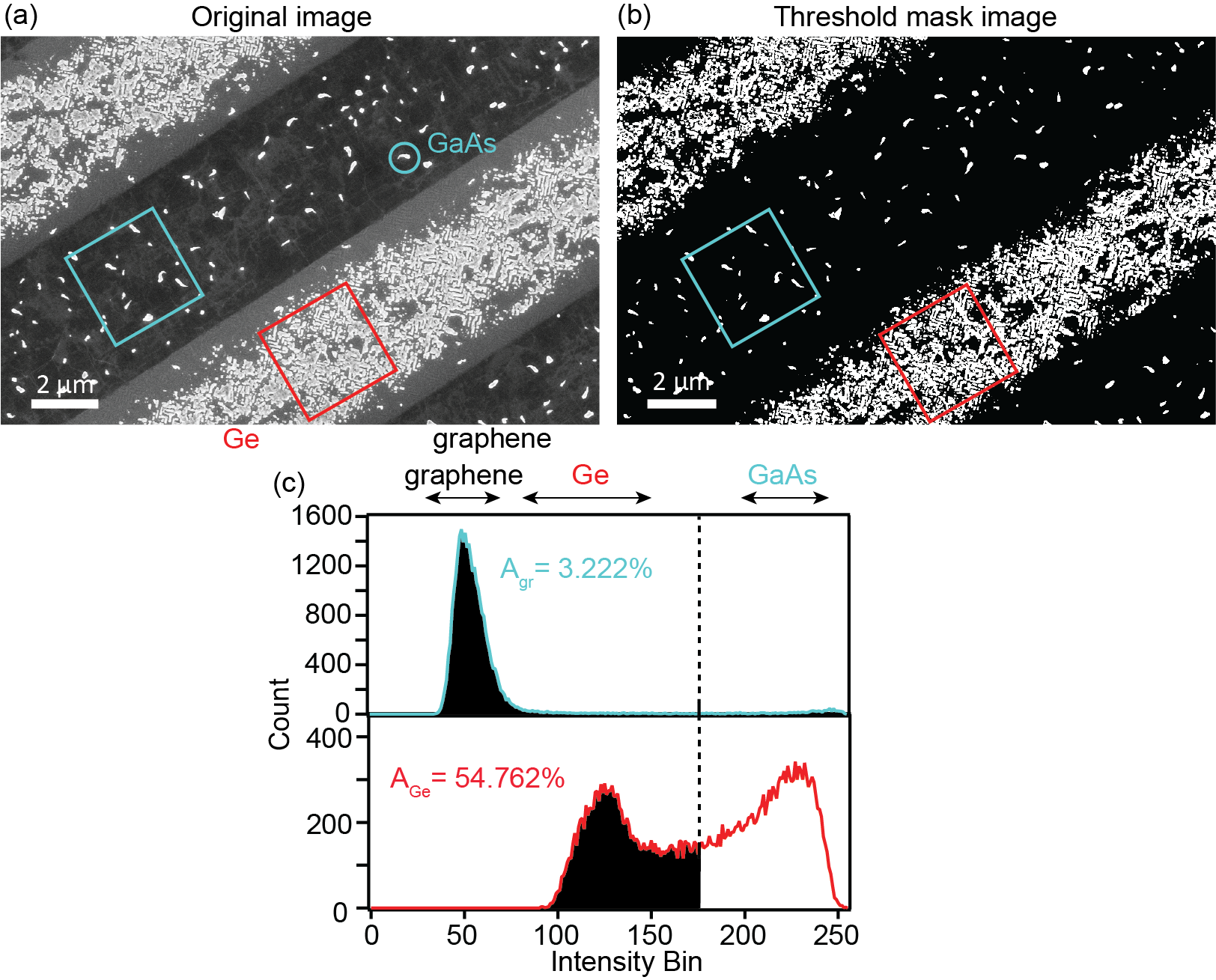}
    \caption{\textbf{Determination of surface coverage.} (a) SEM image of a GaAs film grown at $580^\circ$C. (b) Threshold mask image, used to determine the areal coverage of GaAs film. (c) Histograms of the pixel intensity for the red and blue boxed regions. Distinct peaks in the histograms correspond to Ge, graphene, and GaAs. The threshold for the mask is shown as the dashed line: pixels with intensity higher than this threshold correspond to GaAs. The threshold is determined based on the mid-point between the FWHM of graphene and GaAs peaks.}
    \label{supp_coverage}
\end{figure*}

\end{document}